# Mixed atomic-scale electronic configuration as a strategy to avoid cocatalyst utilization in photocatalysis by high-entropy oxides

Jacqueline Hidalgo-Jiménez[1,2], Taner Akbay[3], Xavier Sauvage[4], Tatsumi Ishihara[1,5] and Kaveh Edalati[1,5,*]

[1] WPI, International Institute for Carbon-Neutral Energy Research (WPI-I2CNER), Kyushu University, Fukuoka, Japan
[2] Department of Automotive Science, Kyushu University, Fukuoka, Japan
[3] Materials Science and Nanotechnology Engineering, Yeditepe University, Istanbul, Turkey
[4] Univ Rouen Normandie, INSA Rouen Normandie, CNRS, Groupe de Physique des Matériaux, UMR6634, 76000 Rouen, France
[5] Mitsui Chemicals, Inc. - Carbon Neutral Research Center (MCI-CNRC), Kyushu University, Fukuoka, Japan

To enhance the activity of photocatalysts for hydrogen production and $CO_2$ conversion, noble metal cocatalysts as electron traps and/or acceptors such as platinum or gold are usually utilized. This study hypothesizes that mixing elements with heterogeneous electronic configurations and diverse electronegativities can provide both acceptor and donor sites of electrons to avoid using cocatalysts. This hypothesis was examined in high-entropy oxides (HEOs), which show high flexibility for atomic-scale compositional changes by keeping their single- or dual-phase structure. A new high-entropy oxide was designed and synthesized by mixing elements with an empty d orbital (titanium, zirconium, niobium and tantalum) and a fully occupied d orbital (gallium). The oxide, synthesized by high-pressure torsion followed by calcination, had two phases (88 wt% orthorhombic (*Pbcn*) and 12 wt% monoclinic (*I2/m*)) with an overall composition of $TiZrNbTaGaO_{10.5}$. It exhibited UV and visible light absorbance with a low bandgap of 2.5 eV, low radiative electron-hole recombination and oxygen vacancy generation due to mixed valences of cations. It successfully acted as a photocatalyst for CO and $CH_4$ production from $CO_2$ conversion and hydrogen production from water splitting without cocatalyst addition. These findings confirm that introducing heterogeneous electronic configurations and electronegativities can be considered as a design criterion to avoid the need to use cocatalysts.
***Keywords*:** high-entropy ceramics; photocatalyst; co-catalyst; lattice defects; high-pressure torsion (HPT)

*Corresponding author (E-mail: kaveh.edalati@kyudai.jp; Tel/Fax: +81 92 802 6744)

# 1. Introduction

The energy demand is rapidly increasing the concentration of $CO_2$ in the environment. To mitigate the harmful effects of energy consumption, new technologies, including hydrogen utilization, must be employed. Hydrogen is a highly dense energy carrier that can eventually replace fossil fuels. Nonetheless, hydrogen production is not necessarily environmentally friendly because the vast majority of hydrogen is produced utilizing fossil fuels. The employment of carbon capture and $CO_2$ storage technologies allows the production of so-called blue hydrogen [1]. However, the capture amount is not enough considering the emitted $CO_2$ [2]. Therefore, green hydrogen production methods such as photocatalysis are still essential to solve the problem.

Photocatalysis allows chemical reactions to occur in the presence of renewable sunlight energy [3]. In this process, the light absorption induces the movements of electrons from the valence band to the conduction band, generating charge carriers that can allocate in the surface and take part in reactions [4]. Two photocatalytic processes have high potential to make hydrogen production and utilization technologies nature-friendly: (1) photocatalytic $CO_2$ conversion and (2) photocatalytic water splitting. Photocatalytic $CO_2$ reduction allows the utilization of $CO_2$ from blue hydrogen generation to produce useful products [5]. On the other hand, the water splitting by photocatalysis implies a green hydrogen generation method considering that it uses only a renewable energy source. For efficient progress of these reactions, the photocatalysts require several features such as high light absorbance, appropriate band structure, abundant active sites, good kinetics for charge carrier mobility, low recombination and high chemical/structural stability [4,5]. Another challenge in photocatalysis is finding new materials that exhibit these characteristics without using precious cocatalysts such as platinum and gold. High-entropy oxides (HEOs) have potential because elements including cocatalysts can be directly included in their lattice and not as a separate phase.

HEOs, similar to high-entropy alloys, are materials whose configurational entropy is higher than $1.5R$ ($R$: gas constant). The high configurational entropy is attributed to a large number of principal elements in equiatomic or close-to-equiatomic fractions [6,7]. It is known that these materials have common characteristics such as (1) crystallinity with high flexibility in tuning chemical composition, (2) single or dual-phase structure with high chemical/structural stability due to their low Gibbs free energy, and (3) heterogeneous distribution of electronic structure due to the co-presence of at least five principal elements and cocktail effect [6]. Since 2020, HEOs have shown to be promising candidates for photocatalytic applications [8] due to their tunability [9–11], inherent lattice defects [11,12], hybridized orbitals [8,13] and multiple active sites due to electronegativity differences [14]. In addition to these characteristics, it was suggested that HEOs have the potential to avoid the utilization of cocatalysts, although their activity without cocatalyst addition is low [15]. One of the main roles of cocatalyst utilization is the improvement of charge transfer and separation which increase the catalytic activity [4,16]. This means that the requirement of a cocatalyst might be related to the nature of the elements utilized. For example, the utilization of platinum and gold as cocatalysts is highly related to their work function and ability to trap electrons due to their high electronegativity [16]. However, elements with $d^0$ and $d^{10}$ electronic configurations and low electronegativity, such as titanium and zinc, can weakly trap electrons [17]. The interaction between d orbital electrons can result in a beneficial electron transfer feature [18,19]; however, the question is how to use this feature to develop photocatalysts that can function without cocatalyst.

Herein, the benefit of HEOs in accommodating various elements in their crystal is used to produce a photocatalyst that does not need any cocatalysts. To do this, the high-entropy oxide TiZrNbTaGaO$_{10.5}$ was synthesized by combining elements with $d^0$ electronic configuration and low electronegativity (titanium, zirconium, niobium and tantalum) and an element with $d^{10}$ electronic configuration and high electronegativity for trapping electrons (gallium). The material exhibited photocatalytic hydrogen production and $CO_2$ conversion without cocatalyst addition. A comparison of the activity of HEO without and with the addition of platinum confirms that the noble metal is not necessary to improve the activity of this material. The implementation of the concept used in this study can expand the understanding of heterogeneous photocatalyst design to produce cost-effective and active catalysts for practical applications.

## 2. Experimental Materials and Procedures
### 2.1. Synthesis

To synthesize the HEO, 55.6 mg of $TiO_2$ anatase powder (Sigma Aldrich, 99.8%), 85.9 mg of $ZrO_2$ (Kojundo, 97%), 92.6 mg of $Nb_2O_5$ (Kojundo, 99%), 154.1 mg of $Ta_2O_5$ (Kojundo, 99.9%) and 65.4 mg of $Ga_2O_3$ (Kojundo, 99.99 %) were mixed to reach a nominal composition $TiZrNbTaGaO_{10.5}$. The binary oxides were mixed with a mortar in acetone for 30 min. Subsequently, 360 mg of the powder mixture was pressed into discs with 1 cm of diameter and processed by severe plastic deformation via high-pressure torsion (HPT) for 6 turns under a 6 GPa pressure. The HPT process, like ball milling, allows a good mixture of oxides on a nanometric scale [20]. The sample obtained by HPT was calcinated in air at 1373 K for 24 h. To improve the mixing of elements, the HPT process was repeated for 3 additional turns and calcinated once more at 1373 K for 24 h.

### 2.2. Characterization

The transformation from the binary oxides to a high entropy oxide was examined by X-ray diffraction (XRD) using Cu Kα radiation. The XRD analysis coupled with the Rietveld method was used to reveal the crystal structure of the final HEO. Moreover, crystallite sizes were estimated from XRD profiles using the Halder-Wagner method. Additional information about the characteristic vibrational modes was obtained by Raman spectroscopy using a 532 nm laser source.

The oxidation state of each element was analyzed by X-ray photoelectron spectroscopy (XPS) using a radiation source of Al Kα.

The homogeneity of the chemical composition and microstructure was examined in two scales. Scanning electron microscopy (SEM) equipped with an energy dispersive X-ray spectroscopy (EDS) under an acceleration voltage of 15 keV allowed the microscopic distribution analysis. The nanometric distribution of elements and microstructure was examined by scanning-transmission electron microscopy (STEM) joined with EDS and transmission electron microscopy (TEM) joined with selected area electron diffraction (SAED) at an acceleration voltage of 200 keV. To make samples for STEM and TEM, the HEO was crushed and dispersed on a 3 mm diameter copper grid covered with carbon film.

Optical properties, such as light absorbance and bandgap were evaluated by UV-vis spectroscopy and the Kubelka-Munk theory, respectively. The radiative recombination of the carriers was examined by photoluminescence spectroscopy using a 325 nm laser source. XPS analysis at low binding energies was used to estimate the valence band top and conduct the band structure analysis in combination with the UV-vis results. The formation of oxygen vacancies, which can affect the optical properties, was examined using electron spin resonance (ESR) utilizing a 9.4688-GHz microwave.

### 2.3. Photocatalytic Tests

The activity of HEO was studied using two different photocatalytic reactions: $CO_2$ conversion and water splitting. The $CO_2$ conversion experiment was conducted in a continuous flow quartz reactor of 858 ml volume. In this reactor, 100 mg of HEO, 4.2 mg of $NaHCO_3$ and 500 ml of deionized water were constantly mixed using a magnetic stirrer with a rotation speed of 420 rpm to ensure the suspension homogeneity. Additionally, the temperature was regulated around 293K using a water chiller. The solution was irradiated using a high-pressure mercury light source with a power of 400 W located in the inner space of the reactor. The light intensity that irradiated the photocatalyst was equivalent to 14 $W/cm^2$. On the top region of the reactor, two pipes were connected: one corresponding to the $CO_2$ input at a flow rate of 30 ml/min, and the second for the output gasses which was connected to two gas chromatographs. The first gas chromatograph (Shimadzu GF-8A) was utilized to analyze the hydrogen and oxygen contents, and the second (GL Science GC-4000 Plus equipped with a GL Science MT 221 methanizer) was utilized to analyze the CO and $CH_4$ concentrations. $CO_2$ was injected for 60 min without light irradiation at the beginning of the test to confirm that there were no air or reaction products in the reactor.

The photocatalytic water-splitting experiments were performed to measure the capability of the sample to produce hydrogen. The photocatalytic activity was tested under three different conditions: (1) 50 mg of sample and 27 ml of deionized water; (2) 50 mg of sample, 27 ml of deionized water and 8 ml of methanol as a sacrificial agent; and (3) 50 mg of sample, 27 ml of deionized water, 8 ml of methanol and 0.25 ml of $H_3PtCl_6 \cdot 6H_2O$ (0.01 M) as a source of 1 wt% platinum cocatalyst. The experiments were performed using an 805 ml reactor coupled to a gas chromatograph Shimadzu GF-8A under a full arc of 300 W UV xenon light with a light intensity of 15 $W/cm^2$. The measurements were conducted for 180 min by testing 3.2 ml of gas phase every 30 min. A blank test was performed before irradiation to confirm no leaks occurred and no reaction products were formed without photocatalysis.

## 3. Results
### 3.1. Crystal structure and microstructure

The evolution of the crystal, examined by XRD in each step of the synthesis process, is shown in Fig. 1(a). In the first step and after mixing the powders, the peaks corresponding to all binary oxides are detected. Subsequently, when applying HPT to the disc, the peaks become broader. The peak broadening is a feature observed in many HPT-processed samples due to the grain size reduction and the formation of defects [20]. As the HPT improves the mixing of the binary oxides, it facilitates the interdiffusion during the calcination. Therefore, after the first calcination, the material mainly changed its crystal structure. As many small peaks were still detected, the process was repeated once more to complete the interdiffusion and reactions. Fig. 1(b) shows the final crystal structure observed for the HEO which includes a monoclinic structure and an orthorhombic structure. The crystal structure information, determined by the Rietveld refinement, is shown in Table 1 for the monoclinic ($C2/m$ space group) and orthorhombic (*Pbcn* space group) phases. According to the Rietveld analysis, the fractions of monoclinic and orthorhombic phases are 12 and 88 wt%, and their crystallite sizes are 30 and 25 nm, respectively.

Analysis by Raman spectroscopy is shown in Fig 1(c). The spectra exhibit clear peaks at 123, 264, 421, 602, 665, 782, 919, 978 and 1004 $cm^{-1}$ which correspond to vibrational modes of the two structures. The peak at 264 $cm^{-1}$ can be assigned to O-Nb/Ti bonds, 665 $cm^{-1}$ to $TiO_6$ or $TaO_6$ octahedra, 782 $cm^{-1}$ to $TaO_6$ octahedra, and 919, 978 and 1004 $cm^{-1}$ to $NbO_6$ octahedra of the monoclinic phase [21,22]. These identifications were conducted by considering the structural similitudes of the monoclinic phase with $Ti_2Nb_{10}O_{29}$. On the other hand, the peaks at 421 and 602

cm$^{-1}$, show similitudes with the orthorhombic phase of TiO$_2$ columbite which shares the same space group [23,24].

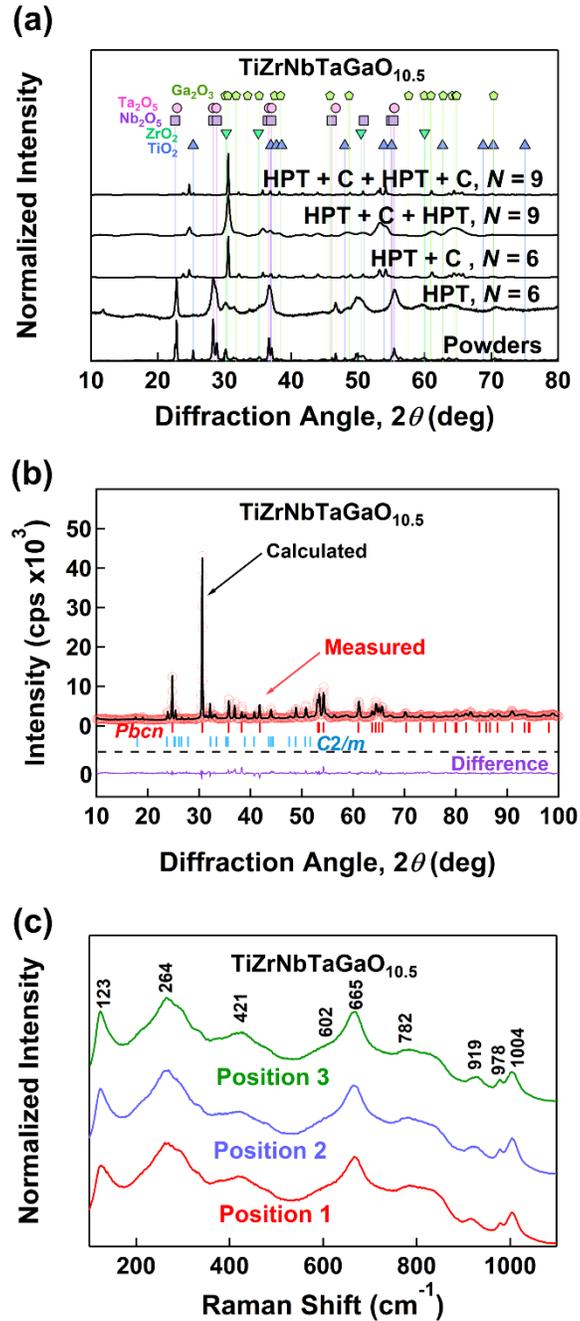

Fig. 1. Development of a dual-phase high-entropy oxide TiZrNbTaGaO$_{10.5}$. (a) Comparison of XRD patterns during the synthesis process, (b) XRD pattern of the final high-entropy oxide and corresponding Rietveld analysis, and (c) Raman spectroscopy obtained in three random positions. C in figure (a) indicates calcination.

Table 1. Lattice parameters of the dual-phase high-entropy oxide TiZrNbTaGaO$_{10.5}$ with monoclinic and orthorhombic crystal structures.

|                      | **Monoclinic**       | **Orthorhombic**   |
|----------------------|----------------------|--------------------|
| Space Group          | *C2/m*               | *Pbcn*             |
| *a* (Å)              | 15.843 ± 0.009       | 4.700 ± 0.004      |
| *b* (Å)              | 3.853 ± 0.002        | 5.569 ± 0.009      |
| *c* (Å)              | 17.946 ± 0.001       | 5.018 ± 0.001      |
| *α* (°)              | 90                   | 90                 |
| *β* (°)              | 102.15 ± 0.03        | 90                 |
| *γ* (°)              | 90                   | 90                 |
| Crystallite Size (nm)| 30 ± 4               | 25 ± 4             |
| Phase Fraction (wt%) | 12 ± 1               | 88 ± 1             |

The oxidation state of the elements present in the HEO is shown in Fig. 2 using XPS analysis. The position of the peaks confirms the complete oxidation of all elements. The peak deconvolution demonstrates the presence of oxidation states of $Ti^{4+}$, $Zr^{4+}$, $Nb^{5+}$, $Ta^{5+}$ and $Ga^{3+}$. The peaks are observed for $Ti^{4+}$ at 458.4 eV and 464 eV, for $Zr^{4+}$ at 181.9 eV and 184.5 eV [25,26], $Nb^{5+}$ at 206.7 eV and 209.5 eV, $Ta^{5+}$ at 25.7 eV and 27.5 eV, and $Ga^{3+}$ at 1118 eV and 1144.2 eV [25,26]. Finally, the analysis for oxygen (Fig. 2(f)) shows a peak at 530 eV. The peaks deconvolution for oxygen show a shoulder to higher binding energies. The presence of this shoulder is an indication of defects such as oxygen vacancies in the material. Point defects, such as vacancies, are commonly observed in high-entropy materials [6]. Moreover, the analysis of the XPS results shows the surface chemical composition, which is 3.1 at% Ti, 2.7 at% Zr, 4.9 at% Nb, 5.2 at% Ta, 9.8 at% Ga and 74.3 at% O. These results suggest that the surface of this material is enriched by gallium atoms.

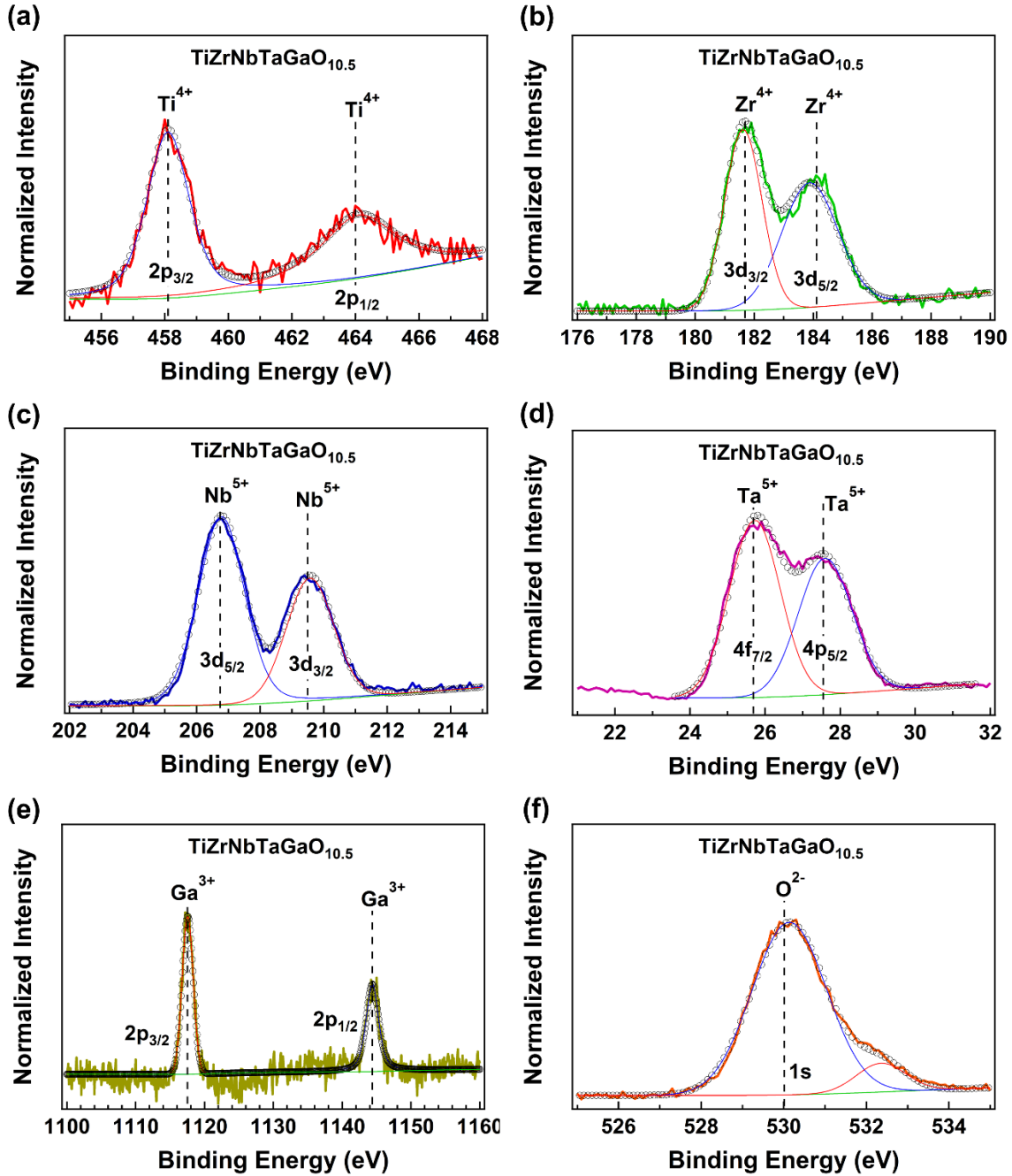

Fig. 2. The oxidation state of elements in the high entropy-oxide TiZrNbTaGaO$_{10.5}$. XPS profiles for (a) titanium, (b) zirconium, (c) niobium, (d) tantalum, (e) gallium and (f) oxygen.

The distribution of elements in the HEO is shown in Fig. 3 using two methods: (a) SEM-EDS and (b) STEM-EDS. SEM-EDS mappings confirm the overall uniformity of the elemental distribution on a microscopic scale. However, some regions rich in tantalum and niobium and poor in titanium, zirconium and gallium appear. Since the fraction of these regions is close to 10%, they should correspond to the monoclinic phase by considering the XRD analysis in Table 1. The quantitative EDS analysis suggests a general composition of 3.2 ± 1.5 at% Ti, 4.0 ± 0.7 at% Zr, 4.7 ± 0.8 at% Nb, 4.6 ± 0.7 at% Ta, 4.9 ± 0.1 at% Ga and 78.7 ± 3.8 at% O. These results are consistent with the nominal composition of TiZrNbTaGaO$_{10.5}$. Further quantitative EDS analysis shows the composition of the secondary phase is 1.3 ± 0.7 at% Ti, 1.9 ± 0.8 at% Zr, 7.3 ± 1.9 at% Nb, 5.8 ± 1.2 at% Ta, 0.7 ± 0.5 at% Ga and 83 ± 3.5 at% O, and the composition of matrix phase is 3.1 ± 1.4 at% Ti, 3.6 ± 0.8at% Zr, 4.0 ± 0.75 at% Nb, 4.0 ± 0.7 at% Ta, 5.0 ± 1.0 at% Ga and 80.3 ± 3.5 at% O. To confirm the crystal structure of the matrix phase and Ta-Nb-rich regions, TEM analysis was conducted on the regions indicated by a square and a circle in Fig. 3(b). Fig. 4. shows the detailed microstructure analysis performed on these regions. Fig. 4(a) and its corresponding SAED in Fig. 4(c) confirm the presence of a monoclinic phase with the 2/*m* space group and [102] orientation in the Ta-Nb-rich region. For the matrix (Fig. 4(b)), SAED analysis (Fig. 4(d)) confirms the presence of the orthorhombic phase with the *Pbcn* space group and [101] orientation. It is concluded that the material has an orthorhombic structure, but small amounts of a Ta-Nb-rich monoclinic phase form in it. Examination of the Ta-Nb-rich region by high-resolution image (Fig. 4(e)) and fast Fourier transform analysis (Fig. 4(f)) also confirm its monoclinic structure. Similarly, high-resolution TEM analysis confirms the orthorhombic crystal structure of the matrix phase, as shown in Fig. 4(g, h). Here, it should be mentioned that these electron microscopy observations indicate larger particle sizes and lower surface area of the HEO compared to starting binary oxide powders (0.01 m$^2$/g versus 9.28 m$^2$/g). This reduction in the surface area can be attributed to the partial consolidation of powder by HPT and calcination [27].

      By considering the crystal structure of the two phases and their chemical compositions, the supercells were modeled via a special quasi-random structure (SQS) configuration using the ICET code with 108 atoms for the orthorhombic phase and 82 atoms for the monoclinic phase. For the orthorhombic phase, an AB$_2$O$_6$-type model with a composition of Ti$_{1/6}$Zr$_{1/6}$Nb$_{2/9}$Ta$_{2/9}$Ga$_{2/9}$O$_6$ was used, in which the A positions are occupied by titanium and zirconium atoms and the B positions are occupied by niobium, tantalum and gallium atoms. For the monoclinic phase, an A$_{12}$O$_{29}$-type model with a composition of Ti$_{1/12}$Zr$_{1/8}$Nb$_{5/12}$Ta$_{1/3}$Ga$_{1/24}$O$_{58}$ was used, in which titanium, zirconium, niobium, tantalum and gallium are randomly distributed. The modeled structures are shown in Fig. 5 for (a) the orthorhombic phase and (b) the monoclinic phase. To confirm the reliability of these models, their XRD patterns were calculated using the powder diffraction utility of the VESTA software and compared with the experimental XRD profiles. As shown in Fig. 5(c), there is a good consistency between the calculated and experimental XRD patterns, confirming the reliability of the modeled structures.

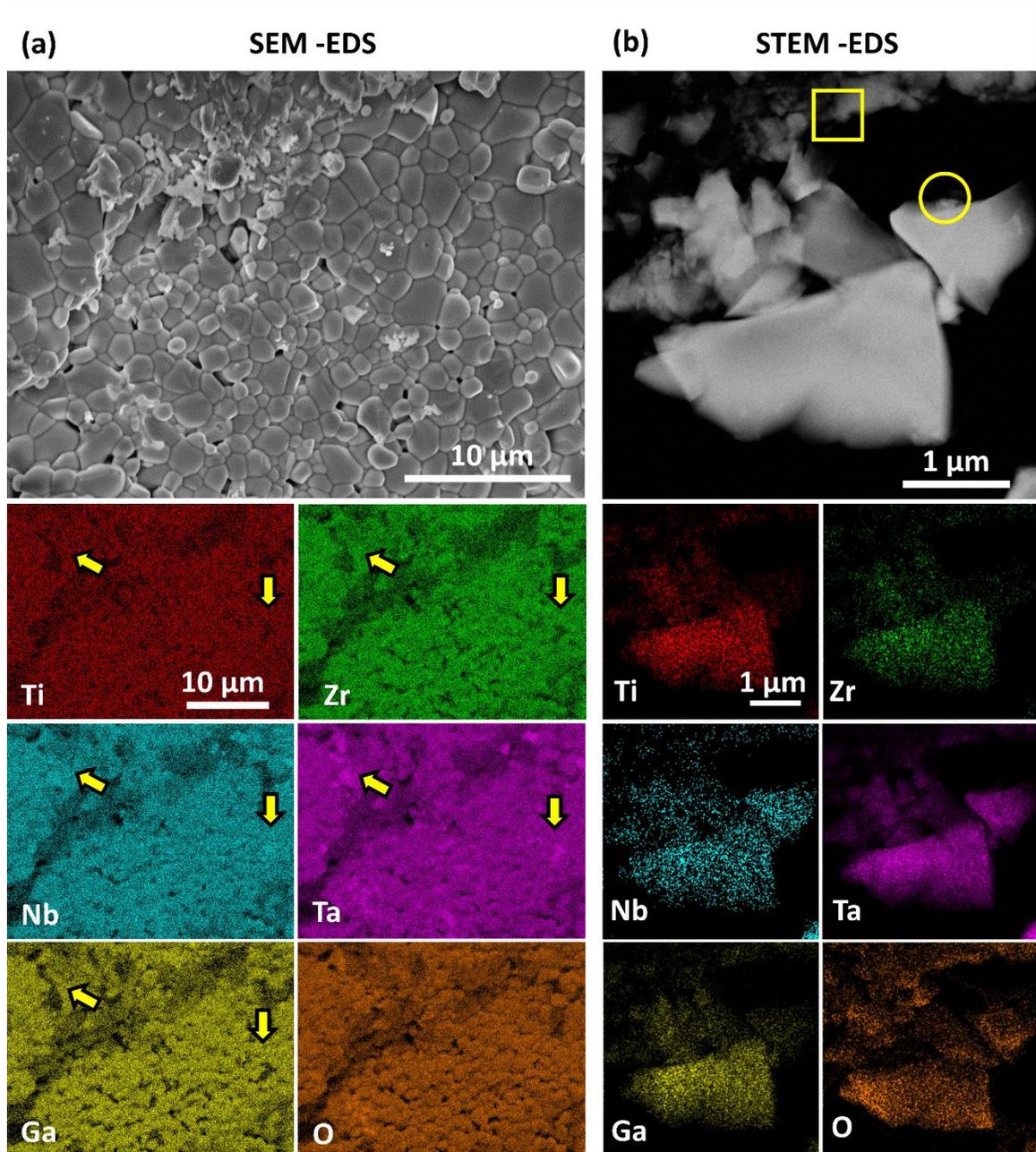

Fig. 3. Elemental distribution of titanium, zirconium, niobium, tantalum, gallium and oxygen in the high entropy oxide TiZrNbTaGaO$_{10.5.}$ (a) SEM back-scatter and (b) STEM high-angle annular dark-field images and corresponding EDS elemental mappings.

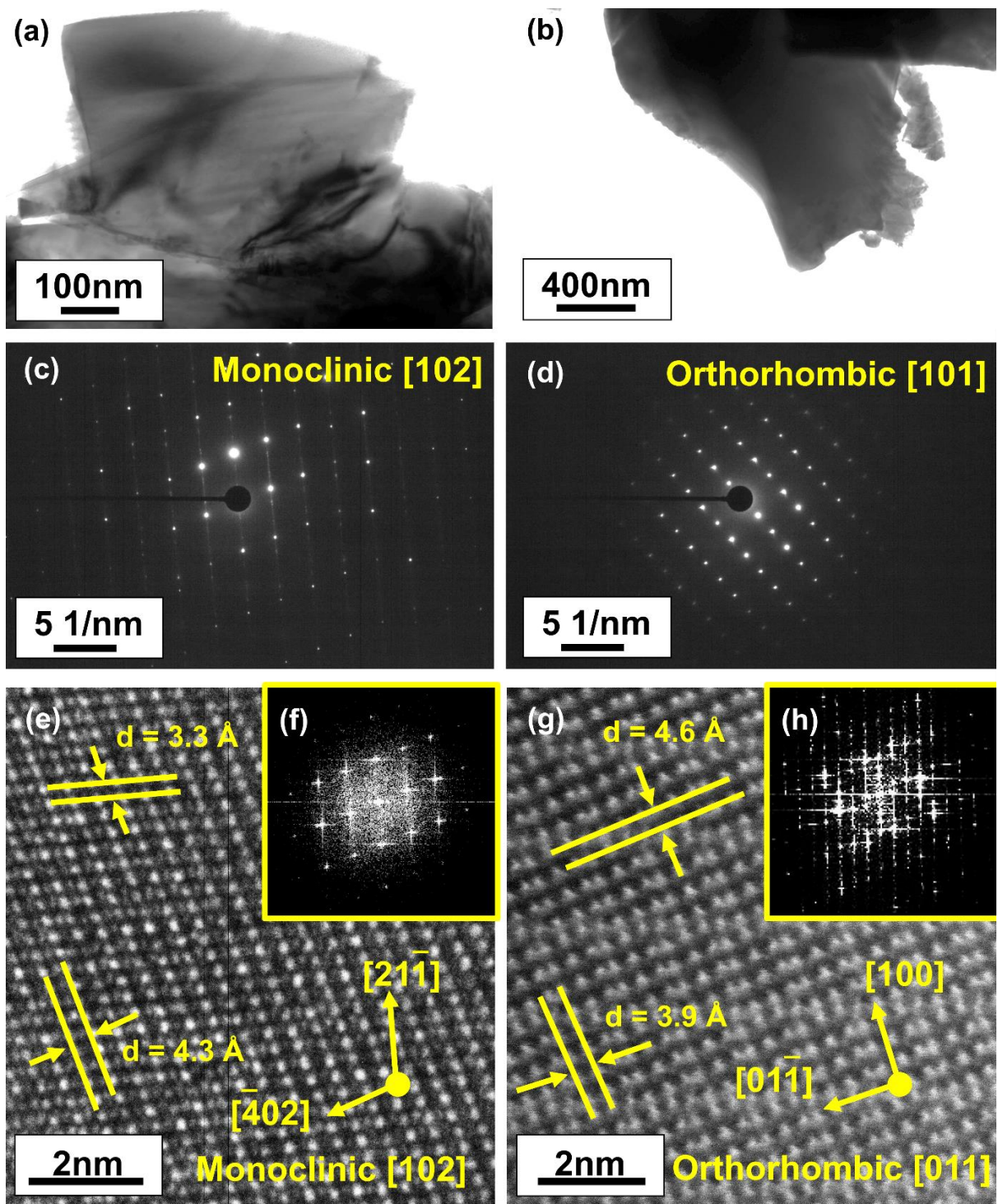

Fig. 4. Microstructural evaluation of the high-entropy oxide TiZrNbTaGaO$_{10.5}$. (a) TEM bright-field image from Ta-Nb-rich region indicated by a circle in Fig. 3(b), (b) TEM bright-field image from matrix indicated by a square in Fig. 3(b), (c, d) SAED profiles corresponding to (a) and (b), (e) high-resolution TEM image of monoclinic phase, (f) fast Fourier transform pattern of (e), (g) high-resolution STEM high-angle annular dark-field image of orthorhombic phase, and (h) fast Fourier transform pattern of (g).

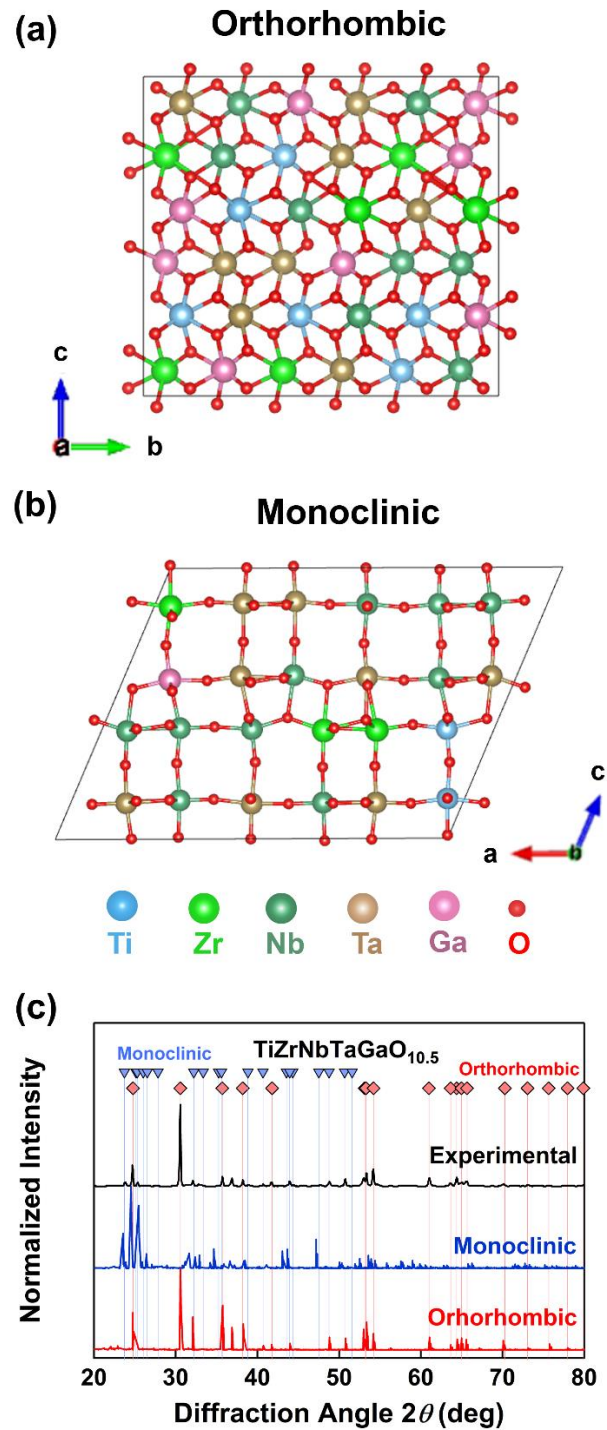

Fig. 5. Structural modeling of the high-entropy oxide TiZrNbTaGaO$_{10.5}$. Modeled structures of (a) orthorhombic phase and (b) monoclinic phase. (c) Comparison of calculated XRD profiles for modeled structures with the experimental profile.

## 3.2. Optical properties

Optical properties examined by UV-vis, XPS and photoluminescence are shown in Fig. 6. Moreover, since oxygen vacancies can affect the optical properties, ESR was also conducted and included in Fig. 6. The light absorbance shown in Fig. 6(a) demonstrates that the HEO can absorb light not only in the UV region but also in the visible light region with significant tail absorbance. The tail absorbance should be due to the presence of color centers such as oxygen vacancies, as evident from the gray color of the sample shown in Fig. 6(a). The UV-vis spectrum was also analyzed by the Kubelka-Munk theory as shown in Fig. 6(b). The bandgap is estimated as 2.5 eV, while another gap is formed close to 1.8 eV. Fig. 6(c) shows the valence band top position obtained by XPS analysis for two points measured independently. The position of the valence band top is approximately 2 eV vs. NHE.

Fig. 6(d) shows the band structure determined by combining the results plotted in Fig. 6(b) and 6(c). The location of the conduction band bottom and the valence band top indicate the suitability of this material as a photocatalyst for both $CO_2$ conversion and water splitting by considering the location of the chemical potentials of reactions such as the oxidation of $H_2O$ to $O_2$ (1.23 eV vs. NHE at pH = 0 or 0.82 eV vs. NHE at pH = 7), the reduction of $H^+$ to $H_2$ (0 eV vs. NHE at pH = 0 or -0.41 vs. NHE at pH = 7) [8,28–31], the transformation of $CO_2$ to CO (-0.11 eV vs. NHE at pH = 0 or -0.52 eV vs. NHE at pH = 7) and the transformation of $CO_2$ to $CH_4$ (0.17 eV vs. NHE at pH = 0 or 0.24 eV vs. NHE at pH = 7) [32,33]. The position of the second gap which should correspond to defect states such as oxygen vacancies is also shown in Fig. 6(d). Here, it should be noted that the measured bandgap in Fig. 6(d) is influenced by the presence of two phases; however, since the orthorhombic phase occupies 88 wt% of the material, it is reasonable to assume that the measured band structure mainly corresponds to this phase.

ESR results, plotted in Fig. 6(e), show the presence of dual peaks with a turning point at a *g* factor of 2.01, suggesting the presence of oxygen vacancies. Due to the lattice distortion by the presence of elements with different atomic sizes and valences, vacancies are common in high-entropy materials [6]. Oxygen vacancies play an important role in photocatalysis: when these defects are located in the bulk they might act as recombination centers, and when they are on the surface they act as active sites for the photocatalytic reaction [34,35]. Fig. 6(f) shows the photoluminescence spectrum of the HEO which has very low intensity. This indicates that the oxide has diminished radiative recombination of the charge carriers even in the presence of oxygen vacancies.

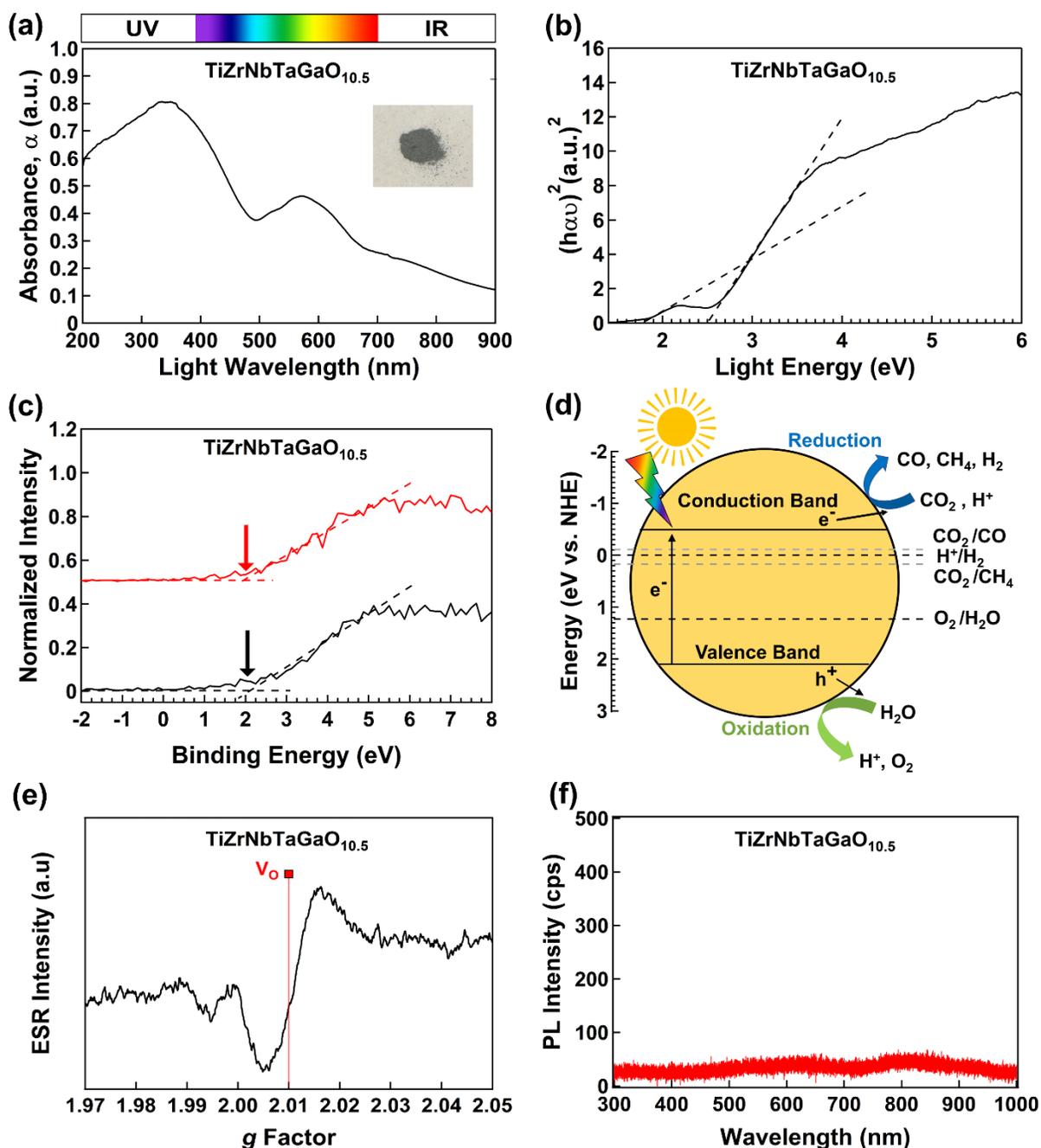

Fig. 6. Appropriate band structure of the high-entropy oxide TiZrNbTaGaO$_{10.5}$ containing oxygen vacancies. (a) Light absorbance by UV-vis spectroscopy, (b) Kubelka-Munk theory analysis for bandgap calculation, (c) XPS valence band top determination obtained at two different positions, (d) band structure and chemical potentials for different water splitting and CO$_2$ conversion reactions, (e) ESR spectrum for oxygen vacancy determination, and (f) photoluminescence spectrum for evaluation of radiative recombination. In (b) $\alpha$ is light absorption, h is Planck's constant and $\nu$ is light frequency.

### 3.3. Photocatalytic activity

Fig. 7(a) shows $CO_2$ to CO conversion and Fig. 7(b) shows the $CO_2$ to $CH_4$ conversion per hour and per surface area of the catalyst. The material produces CO and $CH_4$, but the amount of CO production is higher (average CO and $CH_4$ production rates in 24 h are 14.0 μmol h$^{-1}$g$^{-1}$ and 3.6 μmol h$^{-1}$g$^{-1}$, respectively). There is a gradual decrease in $CO_2$ conversion rate due to the absence of oxidation products such as $O_2$, which results in the accumulation of holes on the surface and reduction of the photocatalytic activity in the absence of a sacrificial agent. To understand the selectivity for $CO_2$ reduction, a comparison using the rate of electron consumption can be utilized [36].

$$\text{Selectivity for CO}_2 \text{ reduction} = [2r(CO) + 8r(CH_4)] / [2r(CO) + 8r(CH_4) + 2r(H_2)] \times 100 \quad (1)$$

where $r$ represents the rate of generation of each product. The selectivity for $CO_2$ reduction for the HEO is 14%, indicating that the developed oxide is more appropriate for water splitting rather than CO2 conversion. For the initial powder mixture of binary oxides, this value is 5%, indicating that it is even more appropriate for water splitting perhaps due to the presence of $TiO_2$. In addition to the $CO_2$ conversion test, water-splitting experiments were also conducted under different conditions, as shown in Fig. 7(c). The three tested conditions were (i) only with water to examine overall water splitting, (ii) with the addition of methanol as a hole scavenger, and (iii) with the addition of methanol and 1 wt% of platinum as a cocatalyst. The sample does not produce any hydrogen or oxygen by using only water. The photocatalytic activity appears by including methanol, but the activity does not change with the addition of cocatalyst. The amount of collected hydrogen indicates that water splitting occurs in the absence of a cocatalyst, although a small part of hydrogen may also come from the transformation of methanol by holes. These results confirm that the HEO already contains elements for trapping electrons (i.e. gallium which is rich on the surface) and it does not require the addition of extra cocatalysts. It should be noted that the photocatalytic activity of the HEO was tested under visible light using a light-emitting diode (LED) lamp with a wavelength of 415 nm; however, no hydrogen was detected during 24 h of irradiation within the detection limits of gas chromatography. Finally, the examination of the HEO by conducting XRD after photocatalytic tests (Fig. 7(d)) indicates its stability for photocatalysis.

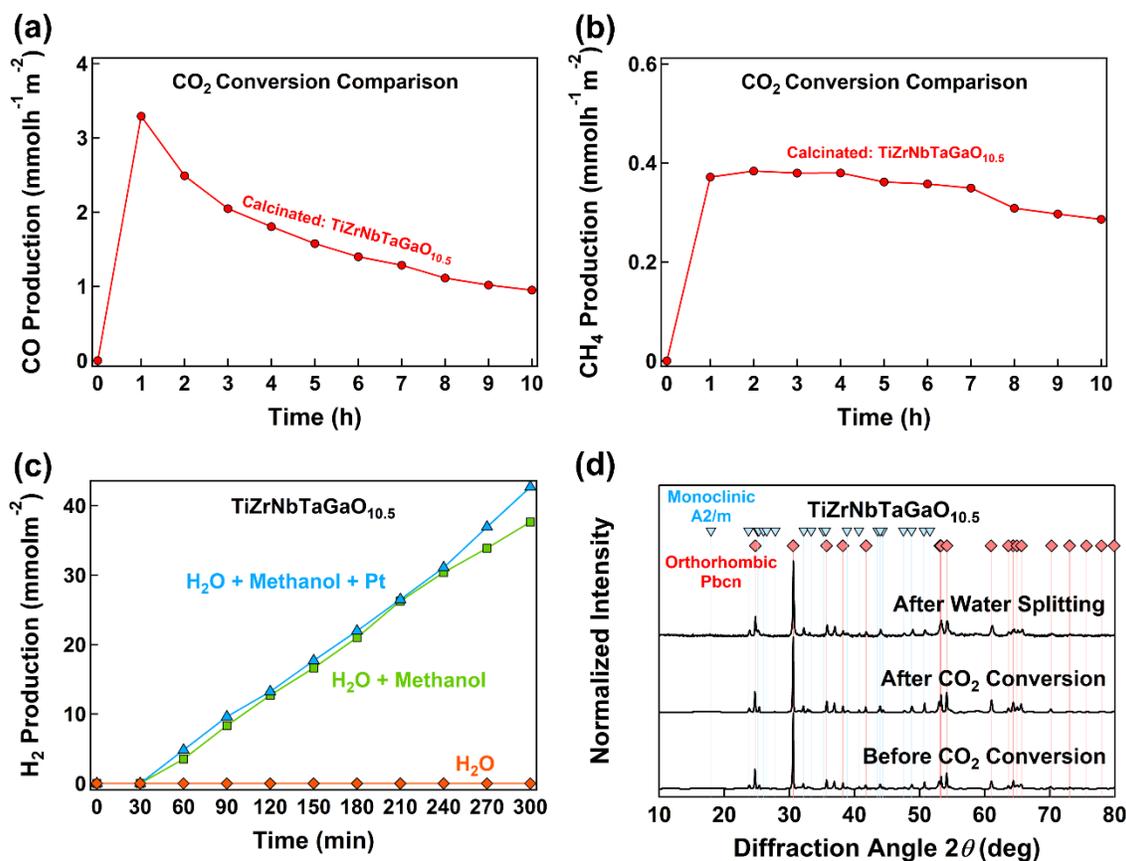

Fig. 7. Photocatalytic activity of the high-entropy oxide TiZrNbTaGaO$_{10.5}$ for CO$_2$ conversion and water splitting without the need for cocatalyst. (a) CO production rate from CO$_2$, (b) CH$_4$ production rate from CO$_2$, (c) H$_2$ production from H$_2$O in three conditions, (i). the presence of only water, (ii) water mixed with methanol and (iii) water mixed with methanol in the presence of platinum cocatalyst, and (d) XRD profiles of oxide before and after catalysis. Production rates and amounts in (a-c) are given per surface area of the catalyst.

## 4. Discussion

A new high-entropy photocatalyst with an overall composition of TiZrNbTaGaO$_{10.5}$ with good photocatalytic activity is introduced in this paper. Contrary to many semiconductors that require the addition of noble metals as cocatalysts, TiZrNbTaGaO$_{10.5}$ shows good activity for CO$_2$ conversion and water splitting without any cocatalyst. Utilizing noble metals makes the photocatalytic methods less cost-effective and thus there is interest in developing catalysts that do not require a cocatalyst. Therefore, the reason why TiZrNbTaGaO$_{10.5}$ does not need any cocatalyst can enhance our understanding of the design of new materials for photocatalytic application. This issue is discussed in the following paragraphs.

For the design of TiZrNbTaGaO$_{10.5}$, elements with different electronic configurations (d$^0$: titanium, zirconium, niobium and tantalum, and d$^{10}$: gallium) were selected. The elements with d$^0$ electronic configuration strongly contribute to the conduction band minimum, work as electron donors and have good performance for photocatalytic applications [37]. The d$^{10}$ electronic configuration elements have the advantage of having conduction bands with high band dispersion due to the contribution of sp orbitals which allow high charge mobility and promote charge separation [38,39]. Gallium with the $d^{10}$ electronic configuration has a high electronegativity and can act as an electron trap site, although its electronegativity is not as high as platinum or gold which are used as cocatalysts for photocatalysis [14]. The presence of gallium on the surface, combined with this feature, improves the adsorption of water molecules. In other words, gallium dispersed on an atomic scale among other elements with lower electronegativities, such as titanium, zirconium, niobium and tantalum, can act as a cocatalyst for photocatalytic hydrogen production. The benefit of combining cations from both sides of the periodic table was also observed in hematite where $Hf^{4+}$ ($d^0$) and $Ga^{3+}$ ($d^{10}$) had a positive impact on the charge transfer and the photocatalytic efficiency of the material [40]. Besides, $Ga^{3+}$ as a dopant in other semiconductors was reported as a good promotor for water oxidation by inhibiting the carrier recombination via forming electron acceptor sites next to electron donor sites [29,41]. By replacing gallium with hafnium as an element with $d^0$ electronic configuration and producing $TiZrHfNbTaO_{11}$ (only $d^0$ elements), not only a cocatalyst is needed but also the photocatalytic reaction rates decrease from 0.096 mmol $h^{-1}g^{-1}$ to 0.036 mmol $h^{-1}g^{-1}$ for hydrogen production, 14.0 μmol $h^{-1}g^{-1}$ to 4.6 μmol $h^{-1}g^{-1}$ for CO generation and 3.6 μmol $h^{-1}g^{-1}$ to 0 μmol $h^{-1}g^{-1}$ for $CH_4$ generation [8,12]. These comparisons indicate that the utilization of elements with heterogeneous oxidation states can also enhance the activity even in the absence of a cocatalyst.

Different oxidation states and ionic sizes of the elements bring severe changes in the lattice such as the formation of defects, as reported in doping scenarios [6,41,42]. In the current HEO, the addition of elements with different oxidation states also induces defects such as oxygen vacancies as observed in the ESR results (Fig 6(e)). Since gallium has the most different oxidation states among the selected cations, its presence should highly contribute to the formation of vacancies. The presence of oxygen vacancies can be associated with larger light absorbance by the formation of defect states in the band structure (Fig 6(d)). On the surface, these defects can also act as reaction sites to adsorb water and $CO_2$ and direct electrons and holes, although the vacancies in the bulk are usually disadvantageous and can decrease the lifetime of charge carriers [35,43]. The concentration of vacancies can also affect the catalytic activity as both experiments and calculations suggested that a high concentration of vacancies over an optimum level can negatively affect the photocatalytic activity by accelerating the electron-hole recombination [43-45]. In the case of $CO_2$ as a weak Lewis acid, it prefers to be adsorbed on an oxygen vacancy site, accept electrons from the adjacent atoms and transform to intermediate products like CO for further hydrogenation [43,46]. The high selectivity of this HEO for $CO_2$ oxidation as well as the formation of $CH_4$ should be partly due to the formation of vacancies. Additionally, since a small amount of a Ta-Nb-rich phase exists in the HEO, it should contain some small fractions of heterojunctions which can be beneficial for charge separation. For such heterojunctions, it is most likely that surface oxygen vacancies behave as an interphase pathway for charge separation and transfer [43]. Taken altogether, mixing cations with different electronic configurations affects the photocatalytic activity both directly by controlling of mobility of the charge carrier and indirectly by the formation of vacancies. Although an earlier study using density functional theory suggested the significance of mixing elements with different electronegativities by severe plastic deformation on the performance of high-entropy photocatalysts [14], future studies are needed to synthesize the nanoparticles of these oxides by methods other than severe plastic deformation and experimentally clarify the mechanisms of high photocatalytic activity of HEOs with mixed $d^0$ and $d^{10}$ electronic configurations.

## 5. Conclusions

A new high entropy photocatalyst was designed by combining elements with different electronic configurations: titanium, zirconium, niobium and tantalum with $d^0$ configuration, and gallium with $d^{10}$ configuration. As a result, $TiZrNbTaGaO_{10.5}$ showed promising properties for photocatalysis such as high light absorbance and good activity for $CO_2$ conversion and water splitting without the addition of any cocatalyst. The combination of elements with different electronic configurations, oxidation states and electronegativities promotes charge mobility and separation by providing electron acceptor sites (i.e. gallium with high electronegativity which acts as a cocatalyst) and electron donor sites. The present study introduces the concept of mixed electronic configurations of oxides as a strategy for designing active photocatalysts that do not need a cocatalyst.

**Acknowledgments**

The authors thank Mr. Fabien Cuvilly of the University of Rouen Normandy for APT sample preparation and analyses. The author JHJ acknowledges a scholarship from the Q-Energy Innovator Fellowship of Kyushu University. This study is supported partly by Mitsui Chemicals, Inc., Japan, partly through Grants-in-Aid from the Japan Society for the Promotion of Science (JP22K18737), partly by Japan Science and Technology Agency (JST), the Establishment of University Fellowships Towards the Creation of Science Technology Innovation (JPMJFS2132), partly by the ASPIRE project of JST (JPMJAP2332), partly by the University of Rouen Normandy through project BQRI MaP-StHy202, and partly by the CNRS Federation IRMA-FR 3095.